
%
%
%
\font\twelverm=cmr12			\font\twelvei=cmmi12
\font\twelvesy=cmsy10 scaled 1200	\font\twelveex=cmex10 scaled 1200
\font\twelvebf=cmbx12			\font\twelvesl=cmsl12
			\font\twelveit=cmti12
	
\skewchar\twelvei='177			\skewchar\twelvesy='60

\def\rm{\fam0\twelverm}          \def\it{\fam\itfam\twelveit}%
  \def\sl{\fam\slfam\twelvesl}     \def\bf{\fam\bffam\twelvebf}%
                   \def\cal{\fam 2}%

  \textfont0=\twelverm   \scriptfont0=\tenrm   \scriptscriptfont0=\sevenrm
  \textfont1=\twelvei    \scriptfont1=\teni    \scriptscriptfont1=\seveni
  \textfont2=\twelvesy   \scriptfont2=\tensy   \scriptscriptfont2=\sevensy
  \textfont3=\twelveex   \scriptfont3=\twelveex  \scriptscriptfont3=\twelveex
  \textfont\itfam=\twelveit
  \textfont\slfam=\twelvesl
  \textfont\bffam=\twelvebf \scriptfont\bffam=\tenbf
  \scriptscriptfont\bffam=\sevenbf
\newcount\numco\numco=0
\def\nono#1$${#1\eqno(\the\numco)$$}
\everydisplay{\global\advance\numco by 1\nono}
\def\cite#1{#1}
\headline={\ifnum\pageno=1\firstheadline\else
\ifodd\pageno\rightheadline \else\leftheadline\fi\fi}
\def\firstheadline{\hfil}
\def\rightheadline{\hfil}
\def\leftheadline{\hfil}
        \footline={\ifnum\pageno=1\firstfootline\else\otherfootline\fi}
\def\firstfootline{\rm\hss\folio\hss}
\def\otherfootline{\hfil}
\font\tenbf=cmbx10
\font\tenrm=cmr10
\font\tenit=cmti10
\font\elevenbf=cmbx10 scaled\magstep 1
\font\elevenrm=cmr10 scaled\magstep 1
 1

\font\ninerm=cmr9

\font\sevenrm=cmr7
\def\part{\partial}
\def\al{\alpha}

\def\be{\beta}
\def\gam{\gamma}

\def\Gam{\Gamma}
\def\ie{{\it i.e.,}\ }

\def\ee{\hbox{e}}
\def\dd{\hbox{d}}
\def\dD{\hbox{D}}
\def\ttt{{2\over3}}
\def\vw{\hbox{vol.(Weyl)}}
\def\eg{{\it e.g.},\ }
\def\sftt{s.f.t.}
\def\hg{{\hat g}}
\def\Del{\Delta}
\def\Gam{\Gamma}
\def\ssc{\scriptscriptstyle}
\def\papers{1}
\def\sft{2}
\def\double{3}
\def\phil{4}
\def\bard{5}
\def\pol{6}
\def\mmdke{7}
\def\david{8}
\def\gl{9}
\def\polo{10}
\def\dotf{11}
\def\witten{12}

\hsize=6.5truein 
\vsize=9.0truein 
\baselineskip=13truept
\rightline{\elevenrm iassns-hep-92-68}
\rightline{\elevenrm McGill/92-49}
\vskip 1truecm
\parindent=3pc
\baselineskip=10pt
\font\tif=cmr10 scaled\magstep3
\centerline{\tif Off-shell String Physics\footnote{${}^*$}{\ninerm %
\baselineskip=11pt Talk by V.P. at the International %
Workshop on String Theory, Quantum Gravity and the Unification of %
Fundamental Interactions, Rome, 21--26 September, 1992}}
\vglue 1.0cm
\centerline{\tenrm ROBERT C. MYERS}
\baselineskip=13pt
\centerline{\tenit Physics Department, McGill University, Ernest
Rutherford Building,}
\baselineskip=12pt
\centerline{\tenit Montr\'eal, Qu\'ebec, H3A 2T8, Canada}
\vglue 0.3cm
\centerline{\tenrm and}
\vglue 0.3cm
\centerline{\tenrm VIPUL PERIWAL}
\centerline{\tenit The Institute for Advanced Study}
\centerline{\tenit Princeton, New Jersey 08540-4920, U.S.A.}
\vglue 0.8cm
\centerline{\tenrm ABSTRACT}
\vglue 0.3cm
{\rightskip=3pc
 \leftskip=3pc
 \tenrm\baselineskip=12pt
 \noindent
Recent advances in non-critical string theory
allow a unique continuation, preserving conformal invariance,
of critical Polyakov string amplitudes to off-shell momenta.
These continuations
possess unusual, apparently stringy, characteristics, which are unlikely to
be reproduced in a string field theory. Thus our results may be an
indication that some fundamentally new formulation, other than string
field theory, will be required to extend our understanding of
critical strings beyond the Polyakov path integral.
Three-point functions are explicitly calculated.
The tree-level effective potential is computed for the tachyon.
\vglue 0.8cm }
\line{\elevenbf 1. Introduction\hfil}
\vglue 0.4cm
\baselineskip=14pt
\elevenrm
\def\pref#1{${}^{#1}$}
Off-shell amplitudes are of great physical interest in string theory,
just as in field theory.  They are essential
for the derivation of effective actions, \eg the derivation
of effective potentials for particles such as the tachyon and the
dilaton, free from the ambiguities of on-shell trivial field redefinitions.
They can be used to derive measures for integrating over moduli
of space-time instantons in string theory, and for the
calculation of hadronic form-factors when one attempts to interpret
certain aspects of quantum chromodynamics in terms of effective string
theories.  Naturally, given the physics involved, off-shell continuations of
string amplitudes have been studied a great deal in the past, beginning
in the early days of dual models. Despite this effort,
off-shell string amplitudes have proven
to possess a remarkable intransigence.
A more extensive discussion of previous investigations
will be presented
elsewhere\pref{\papers}, but we mention three of these approaches
here to put our work in perspective.

Every string field theory naturally defines off-shell amplitudes\pref{\sft}.
These amplitudes have been known for a long time to possess spurious
singularities that can be traced to the fact that \sftt's construct
surfaces from building blocks (propagators and vertices) of fixed
geometries.  One only obtains results invariant under conformal mappings,
or Weyl rescalings of the geometry, when all the external legs are on-shell.
Thus off-shell \sftt\ amplitudes are not geometric, in the sense
that they cannot be associated to a surface independent of the representation
of the conformal structure on the surface.  Naturally enough,
the singularities
in off-shell amplitudes of different \sftt's differ.  Furthermore,
duality is lost in this approach.

Attempts have also been made to compute the Polyakov path integral
on surfaces with finite
boundaries.  The idea here is that specifying the matter configurations
on the boundaries of the surface defines off-shell amplitudes.
As the boundaries shrink to punctures, one expects to see singularities
for boundary conditions corresponding to on-shell states.  While this
approach appears promising, it remains to properly understand
the r\^ole of reparametrizations of the boundaries. These
reparametrizations are not treated in the calculations that
use the theory of doubled Riemann surfaces\pref{\double},
while the remaining calculations have not taken into account the
anomalous Weyl dependence which such reparametrizations
introduce\pref{\phil}.

In the dual model literature, Bardak\c ci, and
Bardak\c ci and Halpern\pref{\bard}, investigated
spontaneous symmetry breaking by summing over tachyon emissions at
zero space-time momentum.  Faced with the problem that such tachyons are not
vertex operators of dimension (1,1), they introduced a fictitious dimension
and let the momentum in this direction be $\pm1.$  These `charged' tachyon
vertex operators were then (1,1) operators.  It will turn out that this
approach is the closest to the approach based on the Polyakov path integral
that we shall pursue.

\vglue 0.6cm
\line{\elevenbf 2. The Weyl measure in the Polyakov functional integral \hfil}
\vglue 0.4cm
Polyakov's derivation of the connection between conformal anomalies
and the critical dimensions of string theories\pref{\pol}
elucidated a multitude of features of string physics,
gleaned piece-meal in pioneering work.
Space-time scattering amplitudes of string excitations are calculated
as correlation functions of vertex operators in a functional integral
over the metric on the string world-sheet, and the space-time
string configurations:
$$
\bigg\langle \prod_i \int\!\dd^2\!z_i \sqrt g \,V_i(z_i)\bigg\rangle
\equiv
\int {\dD g\ \dD X\over \hbox{vol.}(\hbox{Diff})
\hbox{vol.}(\hbox{Weyl})}\ \exp(-S[g,X])
\ \prod_i \int\! \dd^2\!z_i \sqrt g\, V_i(z_i)\ .
$$
The measure is divided by the `volume' of the symmetries of the
classical action $S\equiv (8\pi)^{-1}
\int\!\dd^2\!z \sqrt g g^{ab}\part_a X^\mu\part_b
X_\mu,$ with $\mu=1,\dots,D$---namely, diffeomorphisms and local
Weyl rescalings on the world-sheet. Choosing conformal gauge, $g_{ab}
\equiv\ee^{2\phi}\hg_{ab}(m)$, and fixing diffeomorphisms \`a la
Faddeev-Popov, these functional integrals reduce to
\def\ordre{2}
$$
\int\!\dd m~{\dD \phi\over\hbox{vol.}(\hbox{Weyl})}
{\dD X\ \ \hbox{Det}_{\ssc\rm FP}' \over\hbox{vol.(c.k.v.)}}
\ \exp(-S[\hg,X])
\ \prod_i \int\! \dd^2\!z_i \sqrt {\hat g(m)} V_i(z_i)\ \ .
$$
Here,
c.k.v.~stands for the conformal Killing vectors that must be taken into
account if the world-sheet is a sphere or a torus,
and $\dd m$ denotes the measure for
integrating over moduli labelling distinct conformal equivalence classes
of metrics on surfaces with one or more handles.
In Eq.~(\ordre), the integration
over the Weyl factor should cancel against the volume of the
group of Weyl rescalings in the denominator,
and hence all of the local degrees of freedom in the world-sheet
metric would completely decouple from the theory. This
decoupling is only actually achieved if
Weyl rescaling survives as a symmetry of the quantum path integral.
This requires\pref{\pol} that $D=26$ in order to cancel
the anomalous dependences on the Weyl field, $\phi$, in the
measure factor,
$\dD X\,\hbox{Det}_{\ssc\rm FP}'/\hbox{vol.(c.k.v.)}$.
Also, one must impose various {\it space-time}
 mass-shell and polarization/gauge
conditions on the external string states to avoid any anomalous Weyl
dependences from normal-ordering the vertex operators. Combined these
restrictions ensure that $\phi$ is decoupled
from on-shell correlation functions in critical string theory,
and the Weyl factor simply disappears from the functional integral
(\ie $\int\dD \phi/\hbox{vol.}(\hbox{Weyl}) \equiv 1$).
The presence of the Weyl
volume in the denominator is, of course, justified only by
the fact that we are considering critical (\ie Weyl-invariant) string
theories. For non-critical string theories no such factor occurs.

\def\dtx{\!\dd^2\!x}
\def\dtz{\!\dd^2\!z}
Therefore the mass-shell conditions can be obtained from requiring Weyl
invariance. It follows, in the Polyakov approach, that the
calculation of amplitudes for {\it off-shell}
string states requires the ability to compute correlation functions
of vertex operators with an anomalous Weyl dependence,
in the normalized measure $\dD\phi/\vw .$
Why are such computations difficult? The problem resides in the
non-linearity of the Riemannian metric that defines $\dD\phi.$
The norm on infinitesimal changes of the conformal factor is constructed
with the full world-sheet metric $g_{ab}$
$$
(\delta\phi,\delta\phi) = \int\dtx\sqrt{g}(\delta\phi)^2=
\int\dtx \sqrt{\hat g}\,\ee^{2\phi}(\delta\phi)^2,
$$
which then explicitly depends on $\phi.$  The functional integral over
$\phi$ would be a standard quantum field theory with
the measure, $\dD_{\ssc 0}\phi,$
defined by the translation invariant norm,
$(\delta\phi,\delta\phi)_{\ssc 0} = \int\dtx \sqrt{\hat g}
(\delta\phi)^2.$
The crucial insight that we shall use is due to
Mavromatos and Miramontes, and, independently, D'Hoker and
Kurzepa\pref{\mmdke}.  These authors computed the relation between
these two measures, and found the remarkably simple result
\def\mesure{4}
$$\dD\phi = \dD_{\ssc 0}\phi\ \exp\left(S_{\ssc L} -
{\mu\over \pi}\int \dtz \,\ee^{\al\phi}\right),
$$
where
$S_{\ssc L} \equiv \int {\,\dtz\over6\pi} \bigg[\part\phi\bar\part\phi
+ {1\over4} \sqrt {\hat g} \hat R\phi \bigg]$.
The `cosmological constant' $\mu$ is the coefficient of a local
counterterm, and remains undetermined in this computation.
The constant $\al$ in this interaction is explicitly fixed (see below).
This relation was conjectured originally by David, and Distler and
Kawai\pref{\david}, in their study of two-dimensional gravity coupled
to conformal matter in conformal gauge.  It is important to note that
the derivation of Eq.~(\mesure) is mathematically entirely
independent of the rest of the functional integrals involved.  It is
valid in non-critical string theory, and equally valid in the context of
critical string theory.  It follows, therefore, that any insight into
non-critical string physics, or into quantum Liouville theory, directly
translates into insights into off-shell critical string physics.

\vglue 0.6cm
\line{\elevenbf 3. Correlation functions\hfil}
\vglue 0.4cm
The only assumption in our work is in treating
the correlation functions of interest using the methods of conformal
field theory.  For non-critical strings, this approach has been
verified by comparison with results determined
by matrix model techniques.
The stress tensor deduced from $S_{\ssc L}$ is
$$T_{\ssc L} = {1\over 6} \left[(\part\phi)^2-\part^2\!\phi\right],$$
and it is easily checked that the central charge $c_{\ssc L} = 0.$  Thus
the total central charge for the matter fields, the ghosts and now,
the Liouville field
remains zero.
The weight of an exponential operator $\ee^{\be\phi}$ is
${3\over 2}\be(\be+{1\over 3}).$
An off-shell vertex operator $V_i$ of weight
$(\Del,\Del),$ is dressed in the same way as matter operators in non-critical
string theory to produce a (1,1) operator $\exp(\beta_{\ssc\Del}\phi)V_i$
with
$$\be_{\ssc\Del} = {1\over 6}\left[\sqrt{25-24\Del}-1\right].$$
This is the unique solution of ${3\over 2}\be_{\ssc\Del}
(\be_{\ssc\Del}+{1\over 3})=1-\Del$ such that
$\Del=1\Leftrightarrow \be_{\ssc\Del}=0,$ which insures that in the
on-shell limit, these off-shell amplitudes reduce precisely to
the usual on-shell amplitudes.
Rather puzzling is the non-analyticity in this prescription
at $\Del={25\over 24}.$
While one expects cuts in loop amplitudes in field theories,
it seems difficult to interpret this non-analyticity as arising from
similar physics.
For the present time, we will limit our attention to $\Del\le {25\over 24}$.
Certainly, there is
no obvious physical reason for such a restriction,
and we will address this question further in the concluding remarks.

The presence of the cosmological constant in Eq.~(\mesure)
is important for defining the integration
over $\phi.$ Insertions of cosmological
constant interaction `cancel' Liouville momentum carried by the off-shell
vertex operators, and the background charge term in $S_{\ssc L}.$
However, the treatment of the complete action
is rather subtle\pref{\gl,\polo}. Here, treating the
cosmological constant term as a perturbatively defined interaction,
we determine $\al= \be_{\ssc \Del=0} = \ttt.$
One could consider the other branch of the square root,
which gives $\al=-1, $ but
$\al=\ttt$ may be preferred since then this interaction can be
interpreted as a zero-momentum tachyon, hence as obtained
from the off-shell continuation of a physical state.
Also, if used as the area operator of the quantum
theory, a vanishing area results in the limit
$\phi\rightarrow -\infty,$ in accord with classical expectations.

\def\tg{{3\gam+1\over2}}

Explicit computations can be performed on the two-sphere,
using the idea of Goulian and Li\pref{\gl}
to perform the integral over the constant zero-mode, $\phi_{\ssc 0}.$
The classic calculation of Dotsenko and Fateev\pref{\dotf} can then be used
to compute the resulting
correlation function, with appropriate analytic continuations
along the way\pref{\gl}.  The zero-mode integral is
\def\mess{7}
$$\int d\phi_{\ssc 0} \exp\left({1\over 3}\phi_{\ssc 0}
-{C}{\ee}^{\ttt\phi_{\ssc 0}}
\right) \exp(\gam\phi_{\ssc 0})
= {3\over 2}\Gamma\left({1\over 2}(3\gam+1)\right) {C}^{-{1\over
2}(3\gam+1)},
$$
where $\gam \equiv \sum\beta_i\equiv \sum \be(\Del_i),$
and ${C} \equiv (\mu/\pi)\int \dtz \exp({2\over 3}\tilde\phi),$
with $\int \dtz\,\tilde\phi=0$.
The amplitude is now
$$\Big\langle \prod_i \int \dtz_i\,\ee^{\be_i\phi} V_i(z_i)\Big\rangle
= \hbox{$3\over 2$} \Gamma(-s)
\prod_i\int\dtz_i \Big\langle{C}^s\prod_j \ee^{\be_j
\tilde\phi}(z_j)\Big\rangle_{L}\Big\langle \prod_kV_k(z_k)\Big\rangle_m,
$$
where $s\equiv -{1\over 2}(3\gam+1),$ and the subscript $L(m)$ stands
for Liouville (matter) expectation values.
For three-point functions with positive integer
values of $s$, these correlations were treated by
Dotsenko and Fateev\pref{\dotf}.
Choosing three tachyon operators,
$V_j=\exp(ik_j^\mu X_\mu)$, and
fixing their positions
$\{z_1,z_2,z_3\}$ at $\{0,\infty, 1\}$, yields
$$
{\cal A}=
\hbox{$3\over 2$} \mu^s\Gamma\left(-s\right)\Gamma\left(s+1\right)
\Del({1\over3})^{s}\prod_{k=0}^{s-1}
\prod_{i=0}^{3}\Del(1+2\beta_i+\hbox{$\ttt$} k)\ .
$$
Here $\Del(z) \equiv \Gam(z)/\Gam(1-z),$ and we have defined
$\be_{\ssc 0}\equiv -{1\over 6},$ but $\gam=\sum_{i=1}^3\be_i.$
As it stands this formula is sensible only for positive integer
values of $s$, and hence negative $\gam$.
Using the ideas of Ref.~\cite{\gl},
the above formula can be continued to expressions
which are valid for positive values of $\gam$
$$\eqalign{{\cal A} =
&\left[\mu\Del\left(\hbox{$1\over3$}\right)\right]^{-\tg}
\Gam\left(\hbox{$1+3\gam\over2$}\right)
\Gam\left(\hbox{$1-3\gam\over2$}\right)  \cr & \times
\left(\hbox{$\ttt$}\right)^{\gam-{2\over 3}}
\prod_{i=0}^3\prod^{\gam-\ttt}_{p=0} \Del(1-3\be_i + \hbox{$
3\over 2$}p)\cr
{\rm or}\qquad &\times
\left(\hbox{$\ttt$}\right)^{\chi_+}
\prod_{i=0}^3\prod^{\gam-\ttt(N+1)}_{p=0} \Del(1-3\be_i +\hbox{$
3\over 2$}p)\prod^{N}_{m=1}\Del(1+\gam-2\be_i-\hbox{$\ttt$}m)\cr
{\rm or}\qquad &\times
\left(\hbox{$\ttt$}\right)^{\chi_-}
\prod_{i=0}^3
\prod^{\gam+\ttt(N-1)}_{p=0} \Del(1-3\be_i + \hbox{$
3\over 2$}p)\prod^{N-1}_{m=0}\Del(2\be_i-\gam-\hbox{$\ttt$}m)
,\cr}$$
where $\chi_+=\ttt(4N+1)(N-1)-(4N-1)\gam$ and $\chi_-=
\ttt(4N-1)(N+1)+(4N+1)\gam$. In these formul\ae,
$N$ is a positive integer, and $\gam$ must be such that
the upper limits of the products are integers. This collection of
expressions is clearly redundant, but any two representations
valid for the same value of $\gam$ can be shown to be equivalent.
We illustrate the full set, retaining the auxiliary parameter $N$, since
different formats may prove most useful in examining different problems.
Note that from the first expression and from the last, with $N=1,2$, one
has results which are valid for $\gam=n/3$ where $n$ is a positive
integer or zero. A more extensive description of the analytic continuations
above will appear elsewhere\pref{\papers}.

The amplitudes given above must still be normalized by the division
by the Weyl volume. At tree-level it is possible to
evade a direct computation of the Weyl volume by considering ratios of
amplitudes. It is then interesting to investigate the analytic structure
of the amplitudes
when $N$ and $\gam$ are held fixed.  Considering the ratio of two
 amplitudes (with the same values of $N$ and $\gamma$), one finds
that the interesting dependence on $\be_i$ resides, {\it e.g.},
for $N=1$ and positive integer values of $\gam,$ in
\def\formu{11}
$$
\prod_{i=1}^3 \left\{ \Del(2\be_i-\gam)
\prod^{\gam}_{p=0}\Del(1-3\be_i + \hbox{$3\over 2$}
p)\right\}\ .
$$
This expression is a product of three factors
with poles and {\it zeroes} depending
on the value of $\be_i$ for each individual particle, and $\gam$ as well.
Note that the restriction which arose in the
discussion of the dressings, $\Del\le {25\over 24}$, also constrains
$\beta_i\ge-{1\over6}$. For a fixed $\gamma$, this restricts the
number of poles and zeroes which actually occur.
A case of interest because the
particles can all go on-shell is $\gam=0$, where we find $\prod_{i=1}^3
\Del(1-3\be_i)\Del(2\be_i).$  This expression has poles where
$\beta_i\rightarrow1/3$ (\ie $k^2_i\rightarrow{4\over3}$),
and no zeroes---in particular, it remains finite as
$\be_i\rightarrow 0.$

Explicit computations are possible for other amplitudes. For example,
four point functions are calculable, with
the restriction that one of the particles is either on-shell, and hence
decoupled from the Weyl functional integral, or when one of the
particles is a tachyon at zero momentum, when the amplitude reduces
essentially to the computations above. Amplitudes with any number
of zero-momentum tachyons are readily calculated, as we consider in the
next section.

\def\dxo{\!\dd^{26}\!X_{\ssc 0}}
\vglue 0.6cm
\line{\elevenbf 4. Tachyon Potential \hfil}
\vglue 0.4cm
It is relatively straightforward to calculate an effective
tree-level potential for
the tachyon within our approach. As commented above, the vertex operator for
a zero-momentum tachyon, $\int\dtz\exp({2\over3}\phi)$, is identical to
the cosmological constant interaction. A generating function for connected
tree-level amplitudes for $n$ zero-momentum tachyons is
\def\generous{12}
$$
W(t)=\sum_{n=1}^\infty {t^n\over n!}\bigg\langle
\left[\int \dtz\, \ee^{\ttt\phi} \right]^n \bigg\rangle
=\bigg\langle \exp\left[t\int \dtz\, \ee^{\ttt\phi} \right]-1
\bigg\rangle\ \ .
$$
In these amplitudes,
the operators decouple from the matter part of the functional
integral, and hence the contribution of the latter reduces to
$\int \dxo$. We have set an arbitrary normalization constant to
one, but the zero-mode integral is explicitly retained, as usual.
The exponential of vertex operators shifts
the cosmological constant, $\mu\rightarrow
\mu-\pi t$. In the absence of any vertex operators, Eq.~(\mess)
shows that the unnormalized amplitude is proportional
to $\mu^{-1/2}$, and hence the generating function is
\def\gen{13}
$$
W(t)=\int \dd^{26}X_{\ssc 0}\left(1-{\pi t\over\mu}\right)^{-{1\over2}}\ ,
$$
where we have
dropped the irrelevant constant term in Eq.~(\generous).
This result has a non-analytic singularity at $t=\mu/\pi$ because
on the two-sphere, the $\phi_{\ssc 0}$ integral only converges
with a positive cosmological constant. Hence Eq.~(\gen)
is valid for a source $t<\mu/\pi$.

A Legendre transformation
$\Gamma(T)=-W(t)+\int\dxo\,Tt$ produces an effective tree-level potential
$$
\Gamma(T)=\int\dxo\left[3\left({T\over T_c}
\right)^{1\over3}-{T\over T_c}\right]\ \ ,
$$
\def\pot{14}
where we have introduced an undetermined scale $T_c$.
This uncertainty arises because the correct normalization of $t$, which
would allow for the precise identification of the Polyakov amplitudes
with space-time Green functions, is unknown. It also accounts for
a constant factor which must multiply the sources in Eq.~(\gen) to
correct for the fact that the Polyakov amplitudes are truncated, while
the Legendre transformation requires a generating function for
connected amplitudes with propagators on the external legs.

Further the
above Legendre transformation produces a valid effective potential if
all other string excitations decouple, as follows. Eq.~(\gen) is
the generating function with arbitrary $t$, but all other
sources set to zero, $W=W(t,J^{(m)}=0)$. We have only made a Legendre
transformation with respect to a single source (\ie $t$), and so
Eq.~(\pot) is the correct effective potential with all other fields
set to zero if $\part W(t,J^{(m)})/\part J^{(n)}|_{J^{(p)}=0}=0$
for all currents $J^{(n)}$. This condition is equivalent to the vanishing
of any amplitudes with any number of zero-momentum tachyons and a single
vertex operator for any other string excitation.
For a vertex corresponding to any state
with nonvanishing momentum, such amplitudes vanish as a result of
momentum conservation (\ie the $X_{\ssc 0}$ integrals).
Similarly any zero-momentum vertex operator at higher mass levels
produces vanishing amplitudes because of the matter
or ghost oscillator contributions.
Some states may also have
$\phi$ oscillator contributions---we have not yet proved that these
all decouple.

Let us consider the potential in Eq.~(\pot). One of the most
striking features is that it is non-analytic at $T=0$. This
non-analyticity appears as a result of the singularity in the
generating function, discussed above. Given the restriction
$t<\mu/\pi$, the potential is valid for positive $T$.
In this range, there is a single extremum at $T=T_c$.
Hence one arrives at the conclusion that the bosonic
string cannot be stabilized by a constant shift of
the tachyon.

Perhaps an even more interesting fact made apparent by these
calculations is that the zero-momentum
dilaton $D$ does not couple as expected to the tachyon.
Previous investigations of critical strings give the
expectation that there is a trilinear $DT^2$ interaction.
Such a coupling would lead to singularities in the
amplitudes in Eq.~(\generous), which are in fact not observed.
In discussing the Legendre transformation, we actually showed
that there are no interactions $DT^n$ for any $n$.
Of course, the original expectations are based on calculations
with on-shell tachyons, and so it is perhaps not too surprising
that they are not fulfilled.

\vglue 0.6cm
\line{\elevenbf 5. Conclusions and prospects \hfil}
\vglue 0.4cm
It has been our aim here to show that the effort expended on
the study of non-critical strings in somewhat unphysical contexts has
important physical consequences in critical string theories.
Any future progess in
non-critical string physics, or in quantum Liouville theory, will
be of use in understanding off-shell critical string physics.  There are
a great many physical questions that become accessible in our approach
to off-shell string physics.  The computational limitations of the conformal
field theory treatment of the Liouville correlators obviously leave much
to be desired, a problem that appears in non-critical string theory as well.
Another technical question, with what may be very interesting physics lurking
underneath, is that of the analytic continuation of our discrete
product formul\ae\ as a function of $\gam.$

The restriction of being able to dress operators with dimension
$\Delta \le {25\over 24}$ may be a computational limitation.  On the
other hand, such a limitation is precisely what prevents us from
attempting to sew our amplitudes together as we would if these
amplitudes were off-shell amplitudes of a field theory.  Note that
modular invariance dictates that if we were able to sew amplitudes
together, we would end up with infinite answers.  Thus, it is tempting
to speculate that these non-analyticities are a reflection
of modular invariance, especially since as a function of
space-time momentum, they vary with the mass level
of the vertex operators being dressed.
Since our computations preserved conformal
invariance, any amplitude we compute should be well-defined on moduli
space, {\it i.e.}, should be independent of the coordinates on moduli
space.  This is, of course, not true of \sftt\ off-shell amplitudes,
which describe a particular cell decomposition of moduli space.

There is no conceptual barrier to the extension of our results to
supersymmetric strings, or to open string theories\pref{\papers}.
Above we have only considered simple exponential dressings, but one can
also find many new (1,1) primary fields with Liouville oscillator
contributions (\eg $\part\phi$) which will couple in amplitudes.
Some of these may account for longitudal polarizations which only couple
off-shell\pref{\papers}.

More subtle is the computation of the Weyl volume.  The presence of
the factor $\vw$ in the denominator of Eq.~(\ordre) is
an important feature which distinguishes our off-shell amplitudes from
those of non-critical string. At tree-level though, one can avoid
a direct computation of this factor by considering ratios of
amplitudes. One could follow the
prescription of Ref.~\cite{\gl} which uses Eq.~(\formu) with $\gam=2$
and $\beta_i=2/3$ to compute a result for the two-sphere
(which actually vanishes).
On higher genus surfaces, this factor in the denominator
ensures that the Weyl field does not show up in
any counting of states via degenerations.
In particular, the dependence on the moduli in $\dD \phi$
is precisely cancelled by the
denominator, unless there are off-shell vertex operators present.
Note then that in Eq.~(\ordre), $\dd m$ and $\dD \phi/\vw$ must be
explicitly ordered as given. This crucial cancellation at higher genus
shows that our off-shell continuation is not merely defined by
an additional $c=0$ conformal field theory tacked to a critical string
theory.

A striking feature of the
amplitudes is the presence of poles that are not accounted for by
excitations in the matter sector (even if combined with the ghost sector).
They may indicate the presence of
excitations that are entirely stringy in nature.
Independent of the existence of new poles, the fact that the amplitudes
have products which have upper limits determined by $\gam$ is something
entirely unlike the amplitudes one obtains from a field theory. This
prevents them from factorizing into separate terms depending only
on each individual $\beta_i$. In
field theories, the off-shell character of the amplitude is a function
of individual external states.  Here, one can obtain the value $\gam=0$
when all external states are on-shell, {\it or} if they are off-shell.
It is difficult to imagine how this $\gam$ dependence could be reproduced
in a string field theory. Thus our results may indicate that
some fundamentally new framework, other than string field theory,
will be required to extend our
understanding of critical string theory beyond the Polyakov path integral.

\medskip
Note: Since this talk was given, E. Witten has computed some
off-shell quantities in his background independent
approach to open-string field theory\pref{\witten}.

\vglue 0.8cm
It is a pleasure to
thank the organizers of the Rome Workshop, especially Li\`u
Catena, for their hospitality.
R.C.M. was supported by NSERC of Canada, and Fonds FCAR du Qu\'ebec.
V.P. was supported by D.O.E. grant DE-FG02-90ER40542.

\vglue 0.6cm
\line{\elevenbf References \hfil}
\def\sl{\it}
\def\nuc #1,#2,#3{{\sl Nucl. Phys.} {\bf B#1} (#2) #3}
\def\pr #1,#2,#3{{\sl Phys. Rev.} {\bf #1} (#2) #3}

\vglue 0.4cm
\item{\papers.} R.C. Myers and V. Periwal, Inst. for
Adv. Study/McGill preprints
IASSNS-HEP-92-20/McGill/92-48 (1992);
IASSNS-HEP-92-36 (to appear)
\item{\sft.} J.H. Sloan, \nuc 302,1988,349 ; S. Samuel, \nuc
308,1988,{285, 317};
O. Lechtenfeld and S. Samuel, \nuc 308,1988,{361, {\bf B310} (1988)
254};
V.A. Kostelecky and S. Samuel, {\sl Phys. Lett.} {\bf 207B} (1988) 169;
R. Bluhm and S. Samuel, \nuc 323,1989,{337, {\bf B325} (1989) 275};
B. Sathiapalan, {\sl Phys. Lett.} {\bf 206B} (1988) 211;
D.Z. Freedman, S.B. Giddings, J. Shapiro and C.B. Thorn, \nuc
298,1988,253;
A. Ukegawa, {\sl Phys. Lett.} {\bf 261B} (1991) 391;
J. Feng, \nuc 338,1990,459; K. Sakai, {\sl Prog.  Theor.
Phys.} {\bf 80} (1988) 294
\item{\double.} M. Bershadsky, {\sl Sov. J. Nucl. Phys.} {\bf 45} (1987)
925;
S.K. Blau, M. Clements, S. Della Pietra, S. Carlip and V. Della Pietra,
\nuc 301,1988,285;
J. Bolte and F. Steiner, {\sl Nucl. Phys.} {\bf B361} (1991) 451
\item{\phil.} A. Cohen, G. Moore, P. Nelson and J. Polchinski, {\sl
Nucl.
Phys.} {\bf B267} (1986) 143, {\bf B281} (1987) 127, {\sl Phys. Lett.}
{\bf 169B} (1986) 47;
F. Fucito, {\sl Phys. Lett.} {\bf 193B} (1987) 233;
W.I. Weisberger, {\sl Nucl. Phys.} {\bf B294} (1987) 113;
A.N. Redlich, {\sl Phys. Lett.} {\bf 205B} (1988) 295;
C. Varughese, SUNY (Stony Brook) thesis (1989), UMI-90-11440-mc;
Z. Jaskolski, {\sl Comm. Math. Phys.} {\bf 139} (1991) 353;
M.A. Martin-Delgado and J. Ramirez Mittelbrunn, {\sl Int. J. Mod. Phys.}
{\bf A6} (1991) 1719
\item{\bard.} K. Bardak\c ci, {\sl Nucl. Phys.} {\bf B68} (1974) 331,
{\bf B70} (1974) 397;
K. Bardak\c ci and M.B. Halpern,
{\sl Phys. Rev.} {\bf D10} (1974) 4230,
{\sl Nucl. Phys.} {\bf B96} (1975) 285;
K. Bardak\c ci, {\sl Nucl. Phys.} {\bf B133} (1978) 297
\item{\pol.} A.M. Polyakov, {\sl Phys. Lett.} {\bf 103B} (1981) 207,
211
\item{\mmdke.} N.E. Mavromatos and J.L. Miramontes, {\sl Mod. Phys.
Lett.}
{\bf A4} (1989) 1847; E. D'Hoker and P.S. Kurzepa, {\sl Mod. Phys.
Lett.}
{\bf A5} (1990) 1411; E. D'Hoker, {\sl Mod. Phys. Lett.} {\bf A6} (1991)
745
\item{\david.} F. David,{ \sl Mod. Phys. Lett.} {\bf A3} (1988) 1651;
J. Distler and H. Kawai, {\sl Nucl. Phys.} {\bf B321} (1989) 509
\item{\gl.} M.~Goulian and M.~Li, {\sl Phys. Rev. Lett.} {\bf 66} (1991)
2051; see also A. Gupta, S.P. Trivedi and M.B. Wise, \nuc 340,1990,475
\item{\polo.} A.M. Polyakov, {\sl Mod. Phys. Lett.} {\bf A6} (1991) 635
\item{\dotf.} Vl.S.~Dotsenko and V.~Fateev, {\sl Nucl. Phys.} {\bf B240}
(1984) 312, {\bf B251} (1985) 691
\item{\witten.} E. Witten, Inst. for Adv. Study
preprints IASSNS-HEP-92-53 and -63
(1992)

\vfill
\eject
\bye